\documentclass{ceurart}

\usepackage{listings}
\usepackage{paralist}
\usepackage{tikz}
\usetikzlibrary{arrows.meta,positioning,fit,backgrounds,calc}
\tikzset{
  blk/.style   ={draw=black!55, rounded corners=2pt, line width=0.5pt,
                 align=center, inner sep=3pt, font=\scriptsize,
                 text width=27mm, minimum height=9.5mm, fill=black!3},
  src/.style   ={blk, fill=blue!7,   draw=blue!50},
  mod/.style   ={blk, fill=orange!9, draw=orange!65},
  exp/.style   ={blk, fill=green!9,  draw=green!50!black},
  fl/.style    ={-{Latex[length=1.6mm,width=1.3mm]}, line width=0.5pt, draw=black!60},
  lbl/.style   ={font=\tiny\itshape, inner sep=1.5pt, text=black!60},
}

\begin{document}

\copyrightyear{2026}
\copyrightclause{Copyright for this paper by its authors. Use permitted under Creative Commons License Attribution 4.0 International (CC BY 4.0).}

\conference{4th Italian Workshop on Artificial Intelligence for Cultural Heritage (IAI4CH 2026, https://ai4ch.di.unito.it/), co-located with the 24th International Conference of
the Italian Association for Artificial Intelligence (AIxIA 2026). 6-9 October 2026, Perugia, Italy}

\title{Multimodal Dataset Normalization and Perceptual Validation for Music--Taste Correspondences}

\author[1]{Matteo Spanio}[%
orcid=0000-0002-2436-7208,
email=spanio@dei.unipd.it,
url=https://matteospanio.github.io,
]
\cormark[1]
\fnmark[1]
\address[1]{Department of Information Engineering, University of Padova,
  Via Gradenigo 6/B, 35131 Padova, Italy}

\author[1]{Valentina Frezzato}[]
\fnmark[1]

\author[1]{Antonio Rod\`a}[%
orcid=0000-0001-9921-0590,
email=roda@dei.unipd.it,
url=https://www.dei.unipd.it/~roda/,
]
\fnmark[1]

\cortext[1]{Corresponding author.}
\fntext[1]{These authors contributed equally.}

\begin{abstract}
Music and food traditions are both intangible cultural heritage, and the links between them, how a
sound can make a taste seem sweeter or more bitter, are increasingly used in museum, exhibition and
gastronomic-tourism settings. Modelling those links computationally runs into a data bottleneck
familiar across cultural heritage computing: expert annotation is slow and costly, so the annotated
collections that result are small. The usual remedy is to enlarge a collection automatically,
labelling it with a model trained on the small annotated one. Such \emph{synthetic} labels are rarely
checked, either against the original annotations or against people. We provide both checks.
Experiment~1 asks whether the audio--flavour patterns found in an experimental soundtrack collection
(257~tracks annotated by listeners) survive when the collection is scaled to $\sim$49,300 30-second
segments from the Free Music Archive (FMA) labelled by a fine-tuned Audio Spectrogram Transformer.
Experiment~2 asks whether flavour profiles computed from food chemistry, for 20 dishes drawn largely
from Italian culinary tradition, match what listeners actually hear (49~participants, online).
Feature--flavour patterns carry over for every taste dimension ($\rho=0.38$--$0.72$, all $p<0.001$),
and sweetness still carries over when every spectral feature is removed, so the agreement is not an
artefact of how the labelling model represents audio. Listener ratings match the
computed profiles far beyond chance (permutation $p<0.001$; Mantel $r=0.45$; Procrustes $m^{2}=0.49$),
and the result holds when participants reporting hearing or taste impairments are excluded. We
release the harmonized datasets and all code.
\end{abstract}


\maketitle

\section{Introduction}\label{sec:introduction}
Musical traditions and culinary traditions are both recognized as intangible cultural heritage, and
museums have begun to stage them together. The MUSE science museum in Trento devoted a year-long
exhibition, \emph{Food Sound}, to the ways sound shapes the perception of food, walking visitors in
spatial audio through reconstructed kitchens, markets and dining
rooms;\footnote{\url{https://www.muse.it/en/events/exhibition-food-sound-2025/}, 2025--2026. All URLs
last accessed July 2026.} at the Botanical Garden of
the University of Turin, Vincenzo Guarnieri's installation \emph{Traspirazioni sonore} used designed
sound to evoke the coolness produced by plant
transpiration.\footnote{\url{https://universounito.it/evento/traspirazioni-sonore-installazione-di-vincenzo-guarnieri/},
3--15 June 2025.} Both rest on a mechanism that is by now well documented. People align sensory
dimensions across modalities in consistent ways~\cite{kohler1967gestalt}, and \emph{sonic seasoning}
research shows this extends to taste: sound systematically shifts how flavour is perceived, with
reliable correspondences between taste words and musical attributes such as pitch and
timbre~\cite{crisinel2010bitter,crisinel2010sweet,knoeferle2015sounds},
explained by statistical, semantic and affective
mechanisms~\cite{spence2011crossmodal,motoki2023reflections}, and with measurable effects on
crispness, liking and how long a meal lasts~\cite{dematte2014effects,mathiesen2022sound,galmarini2021impact}.

Exhibitions of this kind are still assembled by hand, one soundtrack at a time. Turning the evidence
into something an institution could apply at scale, a system that proposes music for a dish or
arranges a collection by sensory affinity, requires datasets pairing music with flavour descriptions.
Those datasets run into the constraint that shapes cultural heritage computing generally: the
annotation giving such data its value is made by people, slowly, and therefore in small quantities. Two
limitations follow: scale, because controlled studies cover only a narrow slice of music, and
comparability, because rating scales designed for one study rarely transfer to another.
Recent AI work addresses scale with neural models, metadata fusion and weak supervision, which in
music--flavour contexts buys size at the cost of normalization and validation
problems~\cite{rodriguez-sonic-seasoning-thesis,spanio2025multimodal,fma_dataset}. Feasibility for
music--taste prediction is established~\cite{murari2020key, spanio2026cbmi}, and dataset normalization has been put
forward as the principled response to cross-modal data scarcity, harmonizing heterogeneous sources
into a shared representation instead of collecting more annotations~\cite{spanio2024towards}. No
established protocol checks whether such normalization preserves what matters. We supply one, putting
normalization, transfer diagnostics and perceptual validation in a single workflow.

In this literature a flavour annotation is almost always a \emph{rating}: a participant hears a sound
and scores how strongly it evokes each of a few taste words, typically the five Western basic tastes
(sweet, bitter, sour, salty, spicy), and averaging over participants gives each piece of music a
five-number flavour profile. The instrument is dependable but expensive, so a study yielding a few
hundred annotated stimuli counts as a large one. Our two collections share this format but differ in
where the numbers come from. The \emph{experimental soundtrack collection} carries \emph{human}
annotations, gathered from 22 published studies into 257~tracks; the \emph{FMA-extended corpus}
carries \emph{synthetic} ones, $\sim$49,300 30-second segments from the Free Music Archive
(FMA)~\cite{fma_dataset} labelled on the same five dimensions by an Audio Spectrogram Transformer
(AST)~\cite{gong21b_interspeech} fine-tuned on the smaller
collection~\cite{rodriguez-sonic-seasoning-thesis}.

Enlarging a collection this way buys two orders of magnitude more data, and the practice is now
common wherever heritage collections are enriched automatically. It also raises a question that is
rarely tested: do the patterns that motivated the small collection still hold in the large one, or has
the labelling model reproduced its own biases? We therefore ask whether
audio--flavour relationships carry over from the human-annotated collection to the synthetically
annotated one (\textbf{RQ1}, Experiment~1), and whether flavour profiles computed from food chemistry
and matched to FMA tracks agree with what listeners hear (\textbf{RQ2}, Experiment~2). RQ1 is an
internal check across two sources of labels; RQ2 an external one against newly collected listener
data. Neither settles the question alone, but they constrain it from opposite directions. We
contribute
\begin{inparaenum}[1)]
  \item a transfer analysis showing that audio--flavour correlations, feature rankings and latent
  structure carry over from human-annotated to automatically labelled data;
  \item a perceptual evaluation framework that turns food chemistry into flavour profiles, matches
  them to music and checks the result with listeners;
  \item evidence that the two agree; and
  \item a public release of the harmonized datasets and companion code, reusable for heritage
  applications that pair sound with culinary content.
\end{inparaenum}

\begin{figure*}[tb]
  \centering
  \begin{tikzpicture}[x=1cm, y=1cm]
    \node[src] (sc)   at (0,    2.75) {Soundtrack collection\\257 tracks\\\textbf{human} labels};
    \node[mod] (ast)  at (5.5,  2.75) {Audio Spectrogram\\Transformer\\fine-tuned};
    \node[src] (fma)  at (11.0, 2.75) {FMA corpus\\49{,}300 segments\\\textbf{synthetic} labels};
    \node[mod] (feat) at (5.5,  1.15) {92 shared audio features (\texttt{librosa})};
    \node[exp, text width=52mm] (e1) at (5.5, -0.2)
         {\textbf{Experiment 1}: do the patterns carry over?};

    \draw[fl] (sc)  -- node[lbl, above] {trains} (ast);
    \draw[fl] (ast) -- node[lbl, above] {labels} (fma);
    \draw[fl] (sc.south)  |- (feat.west);
    \draw[fl] (fma.south) |- (feat.east);
    \draw[fl] (feat) -- (e1);

    \node[src] (foodb) at (0,    -1.95) {FooDB\\992 foods\\$\sim$70{,}000 compounds};
    \node[mod] (fart)  at (3.85, -1.95) {FART classifier\\$+$ nutrient map\\(Eq.~\ref{eq:aggregation})};
    \node[mod] (tgt)   at (7.5,  -1.95) {20 food targets\\five taste values\\each};
    \node[exp] (e2)    at (11.0, -1.95) {\textbf{Experiment 2}\\retrieval $+$\\listener study};

    \draw[fl] (foodb) -- (fart);
    \draw[fl] (fart)  -- (tgt);
    \draw[fl] (tgt)   -- (e2);
    \draw[fl] (fma.south) -- node[lbl, right, pos=0.62] {corpus searched} (e2.north);

    \node[font=\scriptsize\bfseries] at (-1.85, 3.5)  {(a)};
    \node[font=\scriptsize\bfseries] at (-1.85, -1.2) {(b)};
  \end{tikzpicture}

  \vspace{1mm}

  \begin{tikzpicture}[x=1cm, y=1cm]
    \node[blk] (p1) at (0,   0) {Consent \&\\demographics};
    \node[blk] (p2) at (3.6, 0) {Instructions:\\five taste terms};
    \node[exp] (p3) at (7.2, 0) {\textbf{10 of 20 tracks}\\random subset\\and order};
    \node[blk] (p4) at (10.8,0) {Debrief};
    \draw[fl] (p1) -- (p2);
    \draw[fl] (p2) -- (p3);
    \draw[fl] (p3) -- (p4);

    \node[mod] (t1) at (3.6, -1.85) {15-s excerpt};
    \node[mod] (t2) at (7.2, -1.85) {rate five tastes\\(7-point scale)};
    \node[mod] (t3) at (10.8,-1.85) {next trial};
    \draw[fl] (t1) -- (t2);
    \draw[fl] (t2) -- (t3);
    \draw[fl] (t3.south) -- ++(0,-0.45) -| (t1.south);

    \draw[densely dashed, draw=black!40, line width=0.4pt] (p3.south west) -- (t1.north west);
    \draw[densely dashed, draw=black!40, line width=0.4pt] (p3.south east) -- (t2.north east);
    \node[lbl, anchor=east, text width=21mm, align=right] at (2.05, -1.85)
         {one trial;\\no food is tasted};
    \node[font=\scriptsize\bfseries] at (-1.85, 0.65) {(c)};
  \end{tikzpicture}
  \caption{Study overview: data sources in blue, processing steps in orange, experiments in green.
  \textbf{(a)} The soundtrack collection trains the labelling model, which in turn labels the much
  larger FMA corpus; both corpora are then described by the same 92 audio features, so the two sets of
  labels can be compared directly. \textbf{(b)} Food chemistry is turned into 20 five-value flavour
  targets, and each target retrieves its closest track from the labelled corpus. \textbf{(c)} Trial
  structure of the listener study: participants only listen and imagine flavours, so no palate
  control is needed.}
  \label{fig:overview}
\end{figure*}

\section{Method}\label{sec:methodology}
Fig.~\ref{fig:overview}a summarizes the workflow. Both experiments share the same five-dimensional
taste space (sweet, bitter, sour, salty, spicy) and the same audio feature space.

\subsection{Data Sources and Harmonization}\label{sec:data}
We jointly analyze three resources: (i) the experimental soundtrack collection, 257~tracks with human
flavour annotations aggregated from 22 published studies~\cite{rodriguez-sonic-seasoning-thesis};
(ii) the FMA-extended dataset, $\sim$49,300 30-second segments from 5,878 FMA
tracks~\cite{fma_dataset} with synthetic annotations; and (iii) FoodDB~\cite{foodb} for target
construction. The FMA labels come from an AST~\cite{gong21b_interspeech} fine-tuned on the 257-track
collection. Experiment~1 therefore tests whether audio-feature--flavour \emph{structure} survives in
model-generated labels, not whether those labels are independently derived, and it inherits whatever
biases the model carries. No FMA audio was used in training and no human-annotated track appears in
the FMA corpus, so there is no direct leakage.

Audio features come from \texttt{librosa}~\cite{mcfee2015} and cover 11 families: zero-crossing rate,
chroma (CQT, STFT, CENS), tonnetz, MFCCs, RMS energy, spectral centroid, bandwidth, contrast and
rolloff, each contributing its mean and standard deviation over the analysis window. The soundtrack
collection yields 148 features and the FMA corpus 92; all cross-corpus comparisons use the 92 the two
share. Analyses that also draw on symbolic descriptors add key and tempo, giving 105 predictors.
Missing values take the per-corpus median.

\subsection{Experiment 1: Cross-Corpus Transfer}\label{sec:exp1}
Experiment~1 compares the two label regimes along three axes, then adds a probe on text metadata.

\emph{Correlation transfer.} For each corpus we correlate every audio feature with every flavour
dimension, then compare the two resulting 92-value profiles per flavour by Spearman rank correlation.
A high value means the features tied to, say, sweetness in one corpus are the same ones tied to
sweetness in the other. We also record sign agreement among the five strongest associations per
dimension, and bootstrap over features to obtain confidence intervals.

\emph{Feature-importance transfer.} For each flavour and corpus we fit Random Forest regressors
(30 bootstrap resamples, 30 trees, maximum depth~8) over the 105 predictors, average the resulting
importances and compare the two rankings by Spearman correlation; FMA is subsampled to 5,000 segments
for tractability. Coverage in the small corpus varies by dimension (from $n=177$ for saltiness down
to $n=58$ for spiciness of 257 tracks), which makes spiciness the least stable. These forests serve
only to rank features, not to predict: they are fit without a held-out split, so we claim no accuracy
for them. As a rough indication of how much flavour variance the audio explains, in-sample partial
least squares reaches $R^{2}=0.38$--$0.50$ on the soundtrack collection and $0.40$--$0.61$ on FMA.

\emph{Latent-structure compatibility.} Horn's parallel analysis counts how many components exceed
chance, and canonical correlation analysis (CCA) measures coupling between the audio and flavour
blocks. CCA is fitted in-sample, so its correlations are optimistic in absolute terms; only the
comparison between corpora is informative.

\emph{Text-side probe.} FMA segments also carry text metadata, so we check whether flavour varies with
genre, mood and free-text description. Genres are mapped to the 16 top-level FMA categories and,
since a track may carry several, expanded to one row per segment--genre pair (89,631 rows); each
flavour is compared across groups by Kruskal--Wallis with Bonferroni-corrected Dunn tests. Mood tags
occurring at least 100 times are compared one-versus-rest with Cohen's~$d$, and free text is
vectorized with TF--IDF and grouped into 8 $k$-means clusters.

\subsection{Experiment 2: Perceptual Evaluation}\label{sec:exp2}
Experiment~2 checks the automatic labels against human perception: we build flavour vectors from food
chemistry, design a target set, match each target to a track and run a listener study.

\subsubsection{From Food Compounds to Taste Vectors}\label{sec:targets}
Compound-level concentrations from FoodDB~\cite{foodb} are converted to taste probabilities by the
FART neural classifier~\cite{Zimmermann2024fart}, a chemical language model mapping a compound's
SMILES string to a distribution over sweet, bitter, sour, umami and undefined; saltiness and
spiciness, which FART does not predict, come from mineral content and pungency descriptors
respectively. Food-level vectors aggregate compounds with logarithmic weighting, following the
Weber--Fechner principle that perceived intensity grows roughly with the logarithm of
concentration~\cite{fechner1860elemente}:
\begin{equation}
W_f(t)= \alpha\sum_i\ln(1+c_i)\,P(t\mid i)  +\beta\sum_j\ln(1+n_j)\,\mathbb{1}[\tau_j=t],
\label{eq:aggregation}
\end{equation}
where $c_i$ is the concentration of compound~$i$ (mg/100\,g), $P(t\mid i)$ its FART-predicted
probability of taste~$t$, $n_j$ the amount of nutrient~$j$ (mg/100\,g) and $\mathbb{1}[\tau_j=t]$
indicates whether nutrient~$j$ maps to taste~$t$. Equation~\ref{eq:aggregation} is our own
formulation, with only the logarithmic compression taken from the literature. The nutrient map covers
37 macronutrients, minerals and vitamins, each carrying a supporting reference from the
sensory-science literature in the released file: carbohydrate is mapped to sweet, sodium and ash to
salty and niacin to sour, while fat, protein and fibre are marked tasteless.

The weights $\alpha$ and $\beta$ set how much the chemistry and nutrient terms each contribute. Only
their ratio matters, since a common factor cancels when vectors are normalized, so we fix $\beta=1$
and choose $\alpha$ by grid search over $[0,3]$ in 16~steps. The criterion rewards weightings that
keep the vectors informative in two ways at once: foods from the same FoodDB group should sit close
together and apart from other groups, measured by the Calinski--Harabasz
index~\cite{calinski1974dendrite}, and each taste dimension should keep a wide range of values rather
than flattening out. Writing $\mathrm{CH}$ for the first quantity and $\bar{s}$ for the mean
per-dimension range, we maximize $\mathrm{CH}+\lambda\bar{s}$ with $\lambda=10$. The selected
$\alpha=0.5$ weights nutrients about twice as heavily as individual compounds. Nutrient amounts are
measured directly; compound contributions are model predictions.

FART outputs six dimensions; the experiment uses five. Umami is folded into saltiness because both
are mediated by dissolved ions and overlap perceptually at low concentrations, and because the
listener study is phrased in the five canonical Western basic-taste terms, which include no common
word for umami. Composite dishes, absent from FoodDB, are built by averaging their ingredient
vectors, and target vectors are per-dimension intensity scores on a 0--100 scale.

\subsubsection{Target-Set Design and Music Matching}\label{sec:targetset}
We use 20 food targets, most of them from Italian culinary tradition: enough to cover the space, few
enough that each still collects a usable number of ratings when every participant hears only half the
set. They mix single-taste items (milk chocolate, radicchio) with multi-taste dishes (ramen, hot and
sour sauce), and were chosen by inspection rather than by a formal coverage criterion. Each target
retrieves the FMA segment closest to it in the five-dimensional space.

\subsubsection{Participants and Procedure}\label{sec:procedure}
Forty-nine participants (17~men, 32~women) completed the study online through
PsyToolkit~\cite{stoet2017psytoolkit}; there was no laboratory session. Ages ranged from 18 to 95
(mean~40.9, $SD=19.2$, median~31), and most reported no formal musical training. Fig.~\ref{fig:overview}c
shows the trial structure: after consenting and answering a demographic questionnaire, participants
read instructions naming the five taste dimensions and then rated a random subset of 10 of the 20
stimuli in random order, with no practice block. Each trial played one 15-second excerpt, taken from
the same segment that was labelled, and asked how strongly the music evoked sweet, salty, sour,
bitter and spicy on 7-point scales. Since the task is auditory and imaginative, no food was presented
at any point and palate cleansing does not arise. Each stimulus received 17 to 32 ratings and the
median session lasted 8 minutes.

Two aspects of the design limit control. Playback hardware was not standardized (35~participants used
device speakers, 13~headphones, one other), and neither ambient noise nor level could be monitored.
Screening recorded, but did not exclude, self-reported sensory impairment: five reported a hearing
impairment and three a taste or smell impairment, seven people in all. Rather than drop these data
silently we report the main analysis on the full sample and repeat it on the unimpaired subsample and
on each playback group (Sec.~\ref{sec:alignment}).

\subsubsection{Statistical Analysis}\label{sec:stats}
Each rating is standardized within its taste dimension and then averaged over the listeners who heard
that track, so every stimulus ends up with one five-number \emph{perceived} profile. Target profiles
are standardized the same way, which puts both on a comparable footing. Two questions then follow.

One is whether \emph{each individual track} was heard the way its own target predicted. We take the
average distance between the 20 paired profiles and compare it against what that average would be if
the pairing carried no information, obtained by repeatedly shuffling which perceived profile goes with
which target (10,000 shuffles). If the real pairing gives a distance smaller than almost every
shuffled one, the match is better than chance.

A different question is whether the two sets have the same overall \emph{shape}, regardless of how
individual items line up: foods far apart in the computed space should also be far apart
perceptually. The Mantel test answers this by correlating the $20\times20$ table of distances between
targets with the same table for the perceived profiles, and Procrustes analysis by sliding, rotating
and rescaling one cloud of 20 points onto the other and measuring the mismatch left over, reported as
$m^{2}$ (0 would mean the clouds coincide) and as the equivalent correlation $r=\sqrt{1-m^{2}}$. Both
are judged against shuffled versions of the data. The two kinds of test answer different things:
item-level accuracy in the first case, the layout of the space in the second. We also report how far
listeners agreed with each other, since that caps how much agreement with any computed profile is
attainable.

A simulation-based power analysis run before data collection, assuming a mixed model
$\text{perceived}\sim\text{target}+(1|\text{subject})+(1|\text{track})$ with a slope of 0.15 and
subject, track and residual standard deviations of 1.2, 1.0 and 1.5, reached the conventional 0.80
threshold at $N=30$; at the realized $N=49$ it is $\approx0.95$. As the analysis reported here works
on track-level averages rather than fitting that model, this speaks to the adequacy of the sample
rather than to the power of the tests themselves.

\section{Results}\label{sec:results}
\subsection{Experiment 1: Cross-Corpus Transfer}
\subsubsection{Audio--Flavour Transfer}
Cross-modal patterns are preserved between corpora. Sign agreement among the top five
feature--flavour associations per dimension is 22/25, so the direction of a relationship is largely
the same whether the labels come from listeners or from the model. First canonical correlations are
0.962 (human-annotated) and 0.910 (FMA); being in-sample they overstate held-out coupling, and the
informative part is their similarity rather than their size.

\begin{figure*}[tb]
  \centering
  \includegraphics[width=\textwidth]{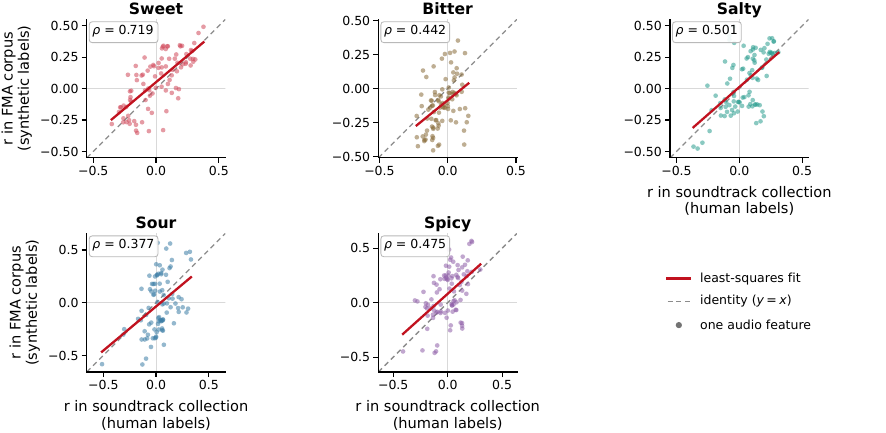}
  \caption{Correlation transfer between corpora, one panel per taste dimension. Each point is one of
  the 92 shared audio features, positioned by its Pearson correlation with that flavour in the
  soundtrack collection (horizontal) and in FMA (vertical); proximity to the dashed identity line
  indicates a preserved association. $\rho$ is the Spearman correlation between profiles.}
  \label{fig:corr-transfer}
\end{figure*}

Fig.~\ref{fig:corr-transfer} shows the transfer directly: each point is one of the 92 shared
features, placed by its correlation with the given flavour in the human-annotated corpus
(horizontal) and in FMA (vertical), so points near the dashed identity line have the same
association in both. All five panels trend positively, sweetness most tightly and sourness most
diffusely. Regression slopes sit below the identity line for most dimensions, so FMA correlations
are compressed towards zero and the labels transfer in direction and rank more faithfully than in
magnitude.

Tab.~\ref{tab:transfer-summary} quantifies this. \emph{Correlation transfer} rank-correlates the two
92-feature profiles, from $\rho=0.377$ (sourness) to $\rho=0.719$ (sweetness), all $p<0.001$, with
bootstrap confidence intervals that stay clear of zero for every dimension. Sweetness carries over
best, consistent with reports of particularly strong cross-modal associations for that
dimension~\cite{crisinel2010sweet}, and sourness worst, plausibly reflecting greater annotation
ambiguity. \emph{Feature-importance transfer} is lower throughout ($\rho=0.328$--$0.516$), as
expected since importance rankings amplify nonlinear and interaction effects sensitive to
corpus-specific variance; bitterness ranks highest, suggesting its predictive cues are the most
stable.

In both corpora the most explanatory feature families are spectral: MFCCs, spectral contrast and
rolloff. That invites an obvious objection, since the labelling model reads spectrograms and might
simply be recycling its own preferred representation. We tested this by discarding every spectral
feature and repeating the analysis on the 52 that remain (chroma, tonnetz, zero-crossing rate, RMS
energy). Sweetness is essentially unaffected ($\rho=0.691$, $p<0.001$) and sourness stays significant
($\rho=0.354$, $p=0.01$), while the other three fall to non-significance
(Tab.~\ref{tab:transfer-summary}). The agreement therefore cannot be put down to spectral
representation alone: for the dimension the literature reports as most robust it survives with the
spectral features gone altogether, though for the remaining dimensions those families do carry most
of the shared signal.

\begin{table}[tb]
  \centering
  {\small
  \begin{tabular}{lccccc}
    \toprule
    & \textbf{Sweet} & \textbf{Bitter} & \textbf{Salty} & \textbf{Sour} & \textbf{Spicy} \\
    \midrule
    Correlation transfer ($\rho$)
      & $0.719^{***}$ & $0.442^{***}$ & $0.501^{***}$ & $0.377^{***}$ & $0.475^{***}$ \\
    \quad 95\% CI
      & \scriptsize[.61,\,.80] & \scriptsize[.23,\,.62] & \scriptsize[.30,\,.67]
      & \scriptsize[.18,\,.54] & \scriptsize[.29,\,.63] \\
    \quad non-spectral features only
      & $0.691^{***}$ & $0.036$ & $0.220$ & $0.354^{*}$ & $0.211$ \\
    \addlinespace[2pt]
    Feature-importance transfer ($\rho$)
      & $0.328^{***}$ & $0.516^{***}$ & $0.448^{***}$ & $0.381^{***}$ & $0.362^{***}$ \\
    \bottomrule
    \multicolumn{6}{l}{\footnotesize $^{*}\,p<0.05$;\; $^{**}\,p<0.01$;\; $^{***}\,p<0.001$.}
  \end{tabular}
  }
  \caption{Agreement between the two corpora, as Spearman rank correlations. \emph{Correlation
  transfer} compares feature--flavour correlation profiles (92 features), with bootstrap confidence
  intervals and a variant computed after discarding all spectral features (52 remaining);
  \emph{feature-importance transfer} compares Random Forest importance rankings (105 predictors).}
  \label{tab:transfer-summary}
\end{table}

\subsubsection{Text--Flavour Associations}
Genre, mood and free-text descriptions carry flavour information too, though weakly. Differences
across the 16 top-level genres are significant for every taste dimension (Kruskal--Wallis, all
$p<0.001$), as are 88--106 of the 120 pairwise contrasts, but these numbers should be read
cautiously: with $\sim$49,300 segments even negligible differences reach significance, group sizes
span two orders of magnitude, segments are nested within tracks, and a segment tagged with several
genres enters several groups at once. Effect sizes are the meaningful quantity and they are modest, with 42
mood tags showing mostly small-to-medium Cohen's~$d$ and TF--IDF clustering yielding only weakly
separated groups (silhouette~$=0.128$). Grouping genres by mean flavour profile nevertheless produces
an interpretable split, with acoustic and traditional genres (Country, Blues, Folk, Pop) leaning
sweet against Hip-Hop, Soul-RnB and Jazz leaning salty and spicy: text--flavour associations are real
but diffuse, and we treat them as corroborating rather than load-bearing.

\subsection{Experiment 2: Perceptual Evaluation}
\subsubsection{Target Construction and Music Matching}
\begin{figure*}[tb]
  \centering
  \includegraphics[width=\textwidth]{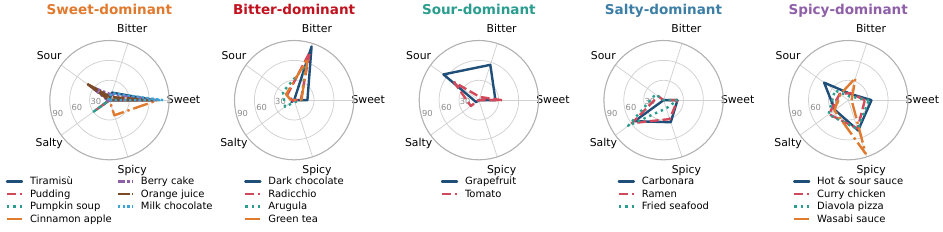}
  \caption{The 20 computational target vectors on a common radial scale (0--90 intensity units).
  Panels group foods by the largest-valued dimension of their target vector; within a panel each food
  has its own colour and line style.}
  \label{fig:target-radar}
\end{figure*}

Fig.~\ref{fig:target-radar} shows the 20 targets grouped by dominant taste, spanning single-peaked
profiles (milk chocolate) and multi-taste dishes (hot and sour sauce), though coverage is uneven:
seven are sweet-dominant and only two sour-dominant. Retrieval quality, scored as
$(1-d/d_{\max})\times100$ with $d_{\max}$ the largest possible
distance~\cite{rodriguez-sonic-seasoning-thesis}, averages $92.2\%$ over a narrow range
($86$--$99\%$). This describes retrieval within the automatic label space only; whether the music
evokes the intended flavour is what the next section tests.

\subsubsection{Perceptual Alignment}\label{sec:alignment}
Listener agreement with \emph{each other} bounds what any model could match, so we report it first.
Individual agreement is modest, as usual for imaginative cross-modal tasks (mean intraclass
correlation 0.27), but averaged over the 17--32 listeners per track the profiles are reliable
(0.89).

All tests agree that those averaged ratings track the computed profiles. The mean distance between
the 20 paired profiles is $d=1.527$, against $d=2.365$ ($SD=0.140$) when the pairing is shuffled;
none of the 10,000 shuffles came as close as the real pairing ($z=5.94$, $p<0.001$). The overall
shape of the space is preserved as well: Mantel $r=0.452$, and Procrustes $m^{2}=0.489$,
equivalently a correlation of $r=0.715$, both $p<0.001$.

These results do not depend on the participants whose inclusion is most debatable
(Tab.~\ref{tab:sensitivity}). Removing the seven who reported a hearing or taste/smell impairment
leaves 42 and makes every statistic slightly \emph{stronger}, and restricting to the 46 participants
under 65 changes nothing. Splitting by playback hardware, the 35 device-speaker users reproduce the
full-sample result closely while the 13 headphone users agree less on structure (Mantel $r=0.308$),
most likely a sample-size artefact since that subgroup leaves only 2--10 ratings per track.

\begin{table}[tb]
  \centering
  {\small
  \begin{tabular}{lccccc}
    \toprule
    \textbf{Subsample} & $N$ & $d$ & \textbf{Mantel} $r$ & $m^{2}$ & \textbf{Procr.} $r$ \\
    \midrule
    Full sample                     & 49 & 1.527 & 0.452 & 0.489 & 0.715 \\
    No reported sensory impairment  & 42 & 1.482 & 0.470 & 0.462 & 0.733 \\
    Age $<65$                       & 46 & 1.528 & 0.446 & 0.490 & 0.714 \\
    Device speakers                 & 35 & 1.492 & 0.438 & 0.482 & 0.720 \\
    Headphones                      & 13 & 1.721 & 0.308 & 0.596 & 0.635 \\
    \bottomrule
  \end{tabular}
  }
  \caption{Sensitivity of the alignment to participant screening and playback hardware. All
  $p$-values are $<0.001$ except Mantel for the headphone subsample ($p<0.01$). Lower $d$ and $m^{2}$,
  and higher $r$, indicate closer alignment.}
  \label{tab:sensitivity}
\end{table}

\begin{figure}[tb]
  \centering
  \includegraphics[width=0.58\columnwidth]{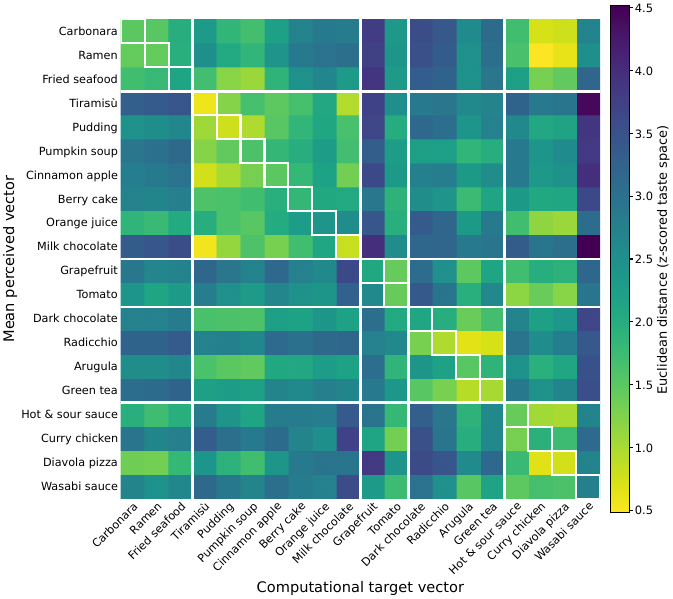}
  \caption{Euclidean distances between mean perceived vectors (rows) and computational targets
  (columns) in the z-scored taste space; brighter is closer. White boxes mark the intended pairs and
  white lines separate dominant-taste groups. Confusions concentrate inside the diagonal blocks.}
  \label{fig:distance-matrix}
\end{figure}

Fig.~\ref{fig:distance-matrix} shows where the agreement succeeds and fails. The intended target is
the closest match for 2 of the 20 tracks and among the three closest for 11, against chance rates of
1 and 3, so the item-level signal is real but modest. The off-diagonal pattern is more informative:
bright cells form blocks along the diagonal, so mix-ups happen almost entirely \emph{within} a
dominant-taste group. Sweet foods are confused with other sweet foods, bitter with bitter, and the
salty and spicy blocks blend together. Listeners recover the dominant taste reliably but cannot tell
apart foods that share it. Single-peaked stimuli such as tiramis\`u accordingly
show the smallest distances and compound profiles such as wasabi sauce the largest, as reported for
cross-modal matching of complex flavours~\cite{crisinel2010sweet}.

This points at how tracks were retrieved rather than at the labels. Euclidean distance treats all five
dimensions as equally important, whereas listeners appear to organize their judgement around whichever
dimension dominates. Cosine distance, which responds to the shape of a profile rather than its
magnitude, bears this out: the intended target becomes the closest match for 5 of 20 tracks instead of
2 and the top-three rate rises from 11 to 13, with agreement still significant ($z=5.31$,
$p<0.001$). Matching the retrieval criterion to how taste judgements are structured would improve
stimulus selection in future studies.

\section{Discussion and Conclusions}\label{sec:discussion}
The two experiments point the same way: audio--flavour structure found in a small, carefully
annotated collection survives automatic scaling, and corresponds to something listeners hear.
Spectral features dominate flavour prediction in both collections, which is easy to dismiss because
the labelling model reads spectrograms, yet removing them entirely leaves sweetness transfer almost
untouched. For the dimension the literature treats as most robust, the agreement therefore reflects
something in the music rather than in the model. The retrieval analysis locates the residual error in
how stimuli were selected rather than in the labels.

\paragraph{Relevance for cultural heritage.} Exhibitions such as those in Sec.~\ref{sec:introduction}
pair sound with food one soundtrack at a time. Scaling that practice, or letting a collection be
browsed by sensory affinity, means working with data that has the difficulty addressed here: the
descriptive layer that gives it value is made by people and therefore scarce, while the raw material
is abundant. Automatic enrichment is the standard response, and our results give an inexpensive
recipe for checking whether it worked: verify that automatic labels preserve the structure of the
human ones, then test that structure against fresh perceptual data. The released collections, pairing
freely licensed music with flavour descriptions and with dishes drawn largely from Italian culinary
tradition, are directly reusable for retrieval or recommendation in such settings. Two caveats apply.
Cross-modal correspondences are shaped by culture, and our participants and dishes are predominantly
Italian, so these profiles are culturally situated rather than universal; extending the protocol to
other traditions would make that cultural component an object of study in itself. And since pairing
sound with food can steer choices, commercial deployments should be transparent about it.

\paragraph{Scope and next steps.} Several limits bound what these results establish. The
labelling model's accuracy is documented in the work that introduced it rather than re-estimated
here, so the transfer analysis is evidence about preserved structure rather than label correctness; a
held-out set of human-annotated tracks would close that gap. The multivariate analyses are fitted
in-sample and describe coupling rather than predict. Stimuli were chosen to lie close to their
targets under the automatic labels, making the perceptual test favourable rather than adversarial, so
a random-selection arm would give a firmer baseline. Finally, the study ran online with
unstandardized playback and its 20 targets were picked by inspection, leaving the sour region thinly
covered. None of these affects the direction of the findings, and because the release spans both
supervision regimes and every intermediate artefact, each can be checked independently.

\begin{acknowledgments}
  Funded by the European Union - NextGenerationEU, under the National Recovery and Resilience Plan
  (PNRR). Code: \url{https://github.com/CSCPadova/music-flavor-analysis}; data:
  \url{https://doi.org/10.5281/zenodo.19259231}.
\end{acknowledgments}

\section*{Declaration on Generative AI}

During the preparation of this work, the authors used the tool Claude Opus 4.8 and Grammarly in order to: Improve writing style, Grammar and spelling check. After using these tools, the authors reviewed and edited the content as needed and take full responsibility for the publication’s content. 

\bibliography{references}

\end{document}